\documentclass[useAMS,usenatbib]{article}
\usepackage{latexsym,amsmath,graphicx}     

\begin{document}
\title{\bf{The Enigma of Star Formation at High Galactic Latitudes}}
\author{Priya Hasan\thanks{e-mail:priya.hasan@gmail.com} \\
Department of Physics\\ Maulana Azad National Urdu University\\
Gachibowli, Hyderabad 500 032 \\}
 
\maketitle
\label{firstpage}
\begin{abstract}
Molecular clouds at very high latitudes ($b > 60^o$) away from the Galactic plane
are considered rare and not conventional sites of star formation. Contrary to this, the recent discovery of high latitude embedded Clusters can possibly change our understanding of the Galaxy formation, evolution and dynamics and the role of the halo in the Galactic evolutionary process. This article reviews a study of nine  embedded clusters (ECs) reported in recent literature with ages less than 5 Myr and vertical distances from the galactic disc ranging from 1.8 to 5 kpc. It discusses the processes that could cause star formation within low density and extraplanar environments in the halo and discuss the possible origins of these clusters. Are these episodic events or is star cluster formation  a systematic phenomenon in the Galactic halo? Two of these objects will be observed by us with Astrosat, so we shall comment on the possible results we expect to get with Astrostat UV and Xray data.

\end{abstract}

\section{Introduction} 

Star Formation is understood as the gravitational collapse of a cool, dense molecular cloud of hydrogen. However, we do not have a very clear understanding of the details of this process, of how young stars are spatially distributed, how mass get distributed and what are the formation processes of low and high mass stars.  Even more enigmatic are the conditions for star formation to determine the kind of stars produced and when does star formation actually occur.

Our galaxy is made up of three components: the disk, halo and bulge.  The halo is made up of an older population of stars that resides in globular clusters. These clusters are made up of low metallicity, denser aggregates of 50,000 to 100,000 stars, gravitationally bound, made up of stellar orbits that are randomly distributed and hence they have spherical shapes. The stars here are redder, older ($\approx$ 10 Gyr) and we do not see any signs of star formation taking place here. The disk component is made up of spiral arms where young stars are constantly forming as it is gas rich and this is where we find young embedded and open star clusters which are looser aggregates of stars with typical lifetimes of a 100 Myr. The nuclear bulge contains the highest density of stars in the galaxy. Although some hot young stars may be found in the nucleus, the primary population of stars there is similar to the old stars found in the halo. The core of the galaxy is obscured by dust and gas at visible wavelengths and can be observed  at other wavelengths with evidence that violent processes may be taking place there. Our galaxy contains a very massive black hole at its center, which drives many of these processes.

This paper deals with recent reports by Camargo et al. (2015, 2016) on the detection of nine embedded clusters (ECs) reported in recent literature with ages less than 5 Myr and vertical distances from the galactic disc ranging from 1.8 to 5 kpc. The discovery of these high latitude Embedded Clusters is fundamental to our understanding of the Galaxy formation, evolution and dynamics and the role of the halo in the Galactic evolutionary process.

\section{Embedded Clusters}
\begin{figure}[h]
\centering
\includegraphics[width=12cm,height=6cm]{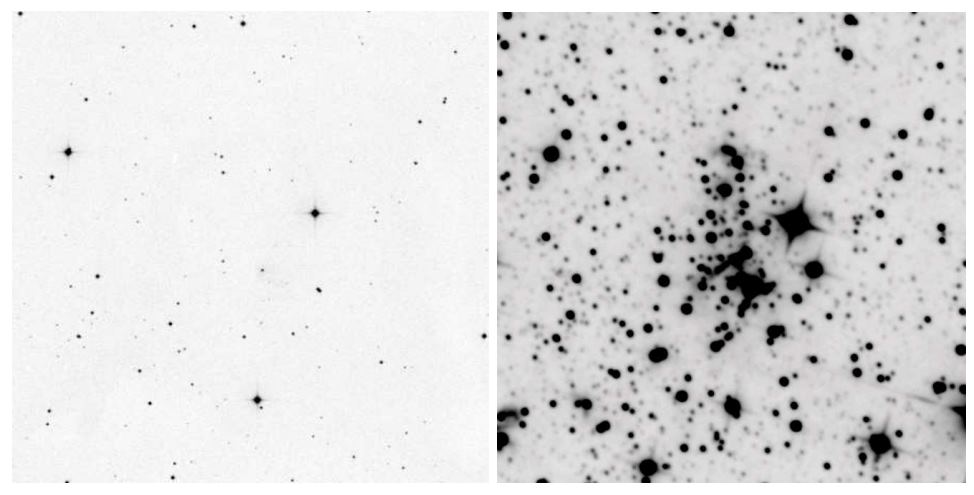}
\caption{An optical (DSS) and infrared (WISE) image of the same field. A cluster of young stars is not apparent in the optical (left) image owing to obscuration by dust, and .  However, a young star cluster is readily apparent in the right image because dust obscuration is significantly less at infrared wavelengths. (image credit: DSS/NASA/IPAC and assembly by D. Majaess).}
\label{oir}
\end{figure}
Embedded clusters (ECs) are clusters that are either completely or partially embedded in gas or dust of ages less than 5 Myr (Lada and Lada, 2003, Hasan, 2016).  ECs have sizes of 0.3-1 pc, and masses 20-1000 M$_{\odot}$, mean stellar   densities $1-10^3$ M$_{\odot}$ pc$^{-3}$. The Star Formation Efficiency is the ratio of the mass of stars to the total mass in a cluster and is  10-30\%. It is this gas that provides the gravitational glue that binds the cluster. These are generally not visible in visible light and can be observed best in infrared wavelengths (Fig. \ref{oir}). The Two Micron All-Sky Survey, or 2MASS, is an astronomical survey of the whole sky in the infrared spectrum (Skrutskie et al. 2006).  Wide-field Infrared Survey Explorer is a NASA infrared-wavelength astronomical space telescope that also provides an all sky coverage in longer wavelengths of 3.4, 4.6, 12 and 22 $\mu$m.

Identification (over density) of young stars can be quantified using Radial Density Profiles (King, 1962) and other methods like the Minimum Spanning Tree and the Nearest Neighbour Method (Casertano \& Hut, 1985). However, detection methods of stars depend on the size, brightness, position in plane, obscuration due to dust, which can vary from cluster to cluster. The WISE bands W1 (3.4$\mu$m) and W2 (4.6$\mu$m) are emitted more by the stellar components while W3 (12$\mu$m)and W4 (22$\mu$m) show dust emission. 
\begin{figure}[h]
\includegraphics[width=10cm,height=5cm]{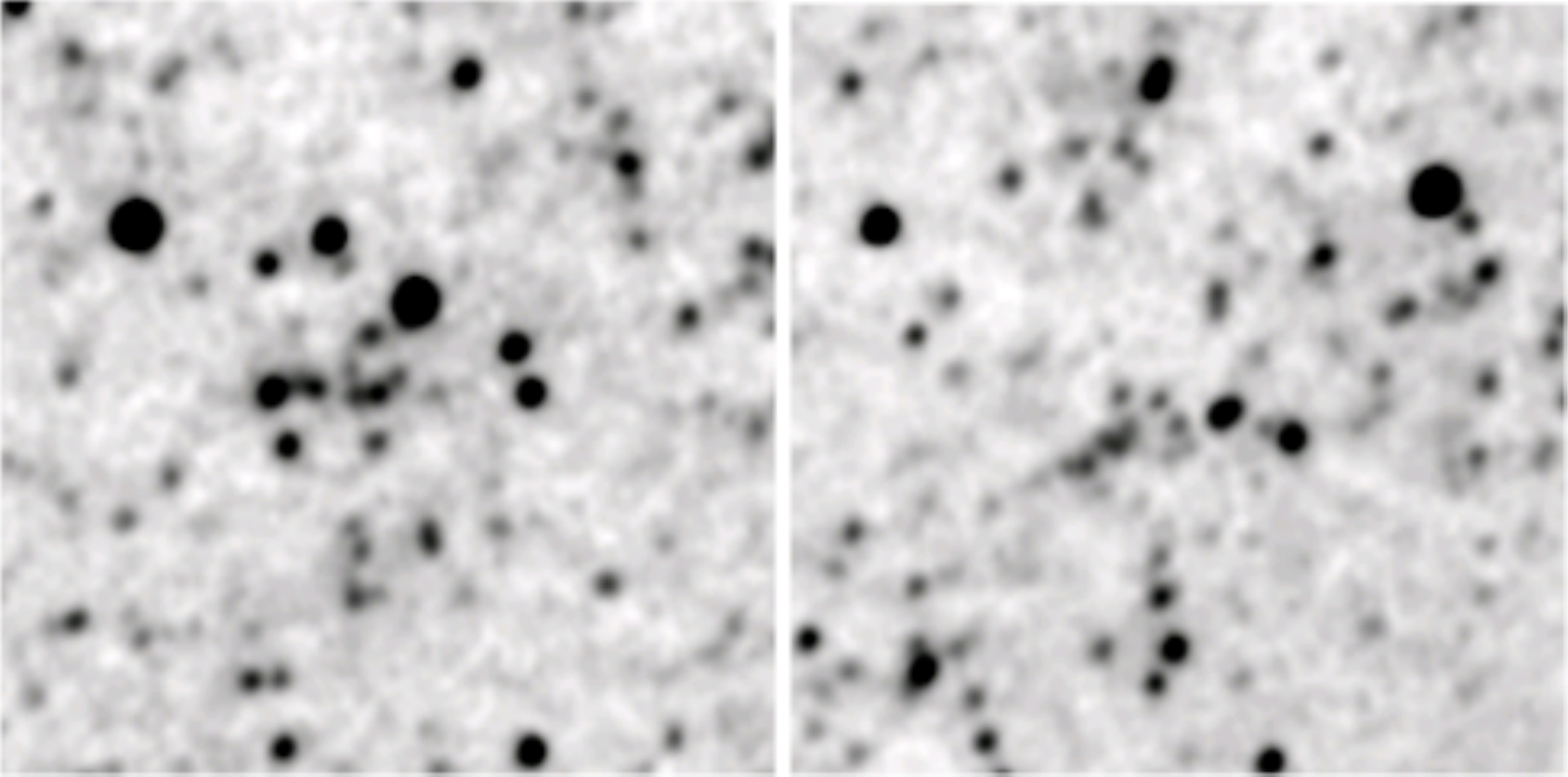}
\caption{The new embedded clusters. Left: WISE W1 ($5′ \times 5′$ ) image centered on the C 439 coordinates. Right: the same for C 438.(Camargo et al.2015)}
\label{mf1893}
\end{figure}

\section{Observations}
\begin{table}[h]
\caption{ Fundamental parameters and Galactocentric components for the ECs (Camargo et al. 2015, 2016)}

\label{sobs}
\small
\begin{center} 
\begin{tabular}{| l |l| l | l |l|l|l|l|}
\hline
Cluster &  $A_V$ &   Age   &  $d_{\odot}$ & $R_{GC}$ & $x_{GC}$  & $y_{GC}$  & $z_{GC}$   \\
        & (mag)  &   (Myr) &  (kpc)       &  (kpc)   &  (kpc)    &  (kpc)    &  (kpc)\\ \hline
C 438  &  0.99 $\pm$ 0.03& 2$\pm$ 1& 5.09 $\pm$ 0.70& 8.69 $\pm$ 0.40& -07.04 $\pm$ 0.02 & +0.97 $\pm$ 0.13& -4.99 $\pm$ 0.69 \\
C 439  &0.99$\pm$ 0.03 &2 $\pm$ 1 &5.09 $\pm$ 0.47& 8.70 $\pm$ 0.26 & -07.05 $\pm$ 0.02& +1.06 $\pm$ 0.10& -4.97 $\pm$ 0.46\\       
C 932   &1.40$\pm$ 0.03& 2 $\pm$ 1&5.7 $\pm$ 0.53&10.55 $\pm$ 0.29&-9.07 $\pm$ 0.17&-0.29 $\pm$ 0.03&-5.38 $\pm$ 0.50\\
C 934&1.46 $\pm$ 0.06&2 $\pm$ 1&5.31 $\pm$ 0.51&10.27 $\pm$ 0.27&-8.97 $\pm$ 0.17&-0.27 $\pm$ 0.03&-5.01 $\pm$ 0.48\\
C 939&1.30 $\pm$ 0.06&3 $\pm$ 2&5.40 $\pm$ 0.50&10.34 $\pm$ 0.27&-9.00 $\pm$ 0.17&-0.31 $\pm$ 0.03&-5.09 $\pm$ 0.47\\
C 1074&0.93 $\pm$ 0.06&3 $\pm$ 1&4.14 $\pm$ 0.39&9.12 $\pm$ 0.15&-8.18 $\pm$ 0.09&-2.66 $\pm$ 0.25&3.02 $\pm$ 0.28 \\
C 1099&0.71 $\pm$ 0.06&5 $\pm$ 1&4.32 $\pm$ 0.61&7.32 $\pm$ 0.30&-6.03 $\pm$ 0.17& -3.61 $\pm$ 0.51&2.05 $\pm$ 0.28 \\
C 1100&0.93 $\pm$ 0.06&1 $\pm$ 1&6.87 $\pm$ 0.36&8.00 $\pm$ 0.23&-4.76 $\pm$ 0.13&-5.59 $\pm$ 0.29&3.16 $\pm$ 0.16 \\
C 1101
&0.96 $\pm$ 0.06&3 $\pm$ 1&3.91 $\pm$ 0.55 &6.83 $\pm$ 0.27&-5.78 $\pm$ 0.20&-3.16 $\pm$ 0.44 &1.78 $\pm$ 0.25 \\    

\hline
\end{tabular}
\end{center}
\end{table}

Camargo et al. (2015) used the WISE data to identify the locations of clusters and then used 2MASS data to find the cluster parameters after careful decontaminations procedures were followed (Table \ref{sobs})\footnote{ $A_V$in the cluster central region,  age, from 2MASS photometry, $R_{GC}$ calculated using $R_{\odot}$ = 8.3 kpc as the distance of the Sun to the Galactic centre,$x_{GC}$, $y_{GC}$, $z_{GC}$: Galactocentric components.}.

\begin{figure}[h]
\includegraphics[width=10cm,height=5cm]{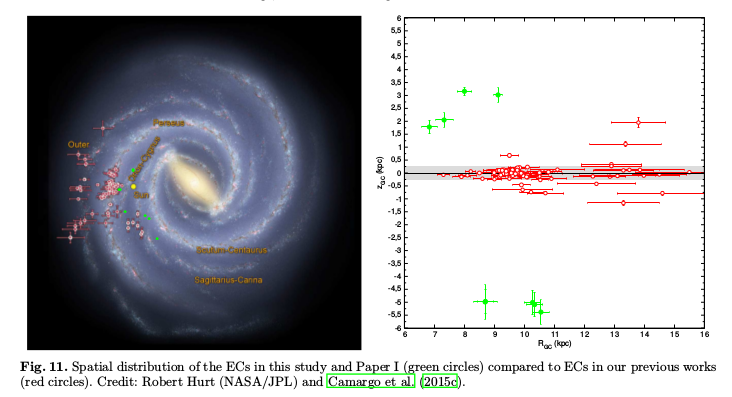}
\caption{Spatial distribution of the ECs in this study (green circles) compared to ECs in  previous works (red circles). Credit: Robert Hurt (NASA/JPL) and Camargo et al.. (2015).}
\label{clus}
\end{figure}

Figure \ref{clus} shows the spatial distribution of the newly found clusters (green circles) compared to the earlier studies )red circles by Carmago et al. (2015, 2016).

\section{Possible Scenarios}
The discovery of these high latitude clusters are very crucial to our understanding of the galactic halo. If these young stars are formed in the halo, then it is possible that these clusters may get unbound before they reach the disc and young stars may just reach the disc isolated. We also need to assess if this is an episodic event or a regular feature. 

There are two possible scenarios that can explain star formation at such high galactic latitudes. One possibile scenario could be Galactic fountains or infall. The accretion of low metallicity gas from the intergalactic medium can happen through filamentary structures with the gas cooling into clouds (Fernandez et al., 2012). C932, C934, and C939 are located right above the Local spiral arm and therefore could relate to the chimney picture. Schlafly et al. (2015) mapped various bubble-like structures vertically along the range 0.3 to 2.8 kpc, which form the Orion superbubble. The expansion of these substructures powered by massive stellar winds and supernovae are triggering star formation
in various shells and rings, inputting energy to the superbubble (Lee \& Chen, 2009). The star formation engine in the Galactic fountain may work as the infall scenario, through the interaction of a cloud with the surrounding halo environment. However, the energetics needs to be worked out to check the feasability of these conditions. Cloud-cloud interactions can also lead to gravitational collapse and triggered star and cluster formation. 

\begin{figure}[h]
\includegraphics[width=10cm,height=5cm]{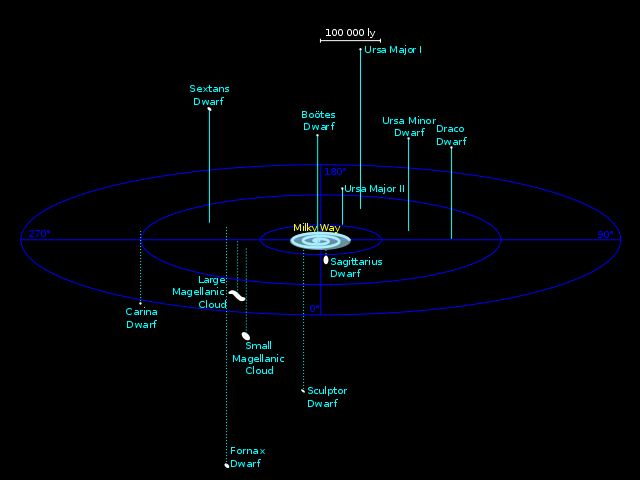}
\caption{Map of the Milky Way and Satellite Galaxies Credit: Creative Commons Attribution-Share Alike 2.5 Generic license}
\label{sat}
\end{figure}

The other possible scenario is extragalactic in nature. Our Milky Way galaxy has several smaller galaxies in its vicinity called `satellite galaxies'. There are 12 known satellites of our galaxy (Fig \ref{sat}). 

Tidal interactions of the galaxy with its satellites is also a possible reason for star formation to take place so far from the disc of the galaxy. Further kinematic information of these stars from GAIS DR2 data could possibly be used to validate such a situation.

\section{Conclusions}
This paper deals with recent reports in literature by Camargo et al. (2015, 2016) on the detection of nine embedded clusters (ECs) with ages less than 5 Myr and vertical distances from the galactic disc ranging from 1.8 to 5 kpc. The discovery of these high latitude Clusters is fundamental to our understanding of the Galaxy formation, evolution and dynamics and the role of the halo in the Galactic evolutionary process. We discuss the possible scenarios in which this can take place. 

Turner et al. (2017) did careful analyses of photometric and star count data available for the above nine young clusters identified by Camargo et al. (2015, 2016) and found that none of the groups contain early-type stars, and most are not significant density enhancements above the field level. This is is absolute contradiction to what was found earlier. It is therefore necessary to carry out systematic observations and/or analysis if existing data to resolve this issue. 

We proposed simultaneous observations  of these clusters using UVIT and the Xray telescopes on Astrosat, the Indian Astronomy Satellite. It is well known that young stars are copious emitters of Xray and so Xrays can also be good tracers of the Young Stellar Populations in these clusters. We can use the Xray Luminosity function to independently derive the distance to these clusters. We can also use Xray spectral data to fit spectra of these sources and derive the column density $n_H$.  
The UV data would be very useful to get a good handle on the extinction of these sources. We shall also use this data to construct SEDs for the YSOs from Xray to ultraviolet and infrared from WISE. We shall also plot color magnitude diagrams for the clusters.  All this information will provide important constrain on theoretical models and are essential to understand the star formation in these clusters at high latitudes.

We hope that, with this data available in ultraviolet and Xray wavelengths, we can further validate these results and exciting insights in star formation in the galaxy will be possible.
GAIA DR2 will  be released in April 2018 and it is hoped that it shall also provide important solutions to this enigma.

\end{document}